\documentclass[useAMS,usenatbib]{mn2e}
\usepackage{txfonts}
\usepackage{graphicx}
\usepackage{textcomp}
\usepackage{natbib}
\usepackage{deluxetable}

\newcommand{\um}{\textmu m }

\def\Dn{D$_n$(4000) }
\def\Dnu{D$_n$(4000)}
\def\Dno{D$_n$(4000)$_0$}

\bibpunct[]{(}{)}{;}{a}{}{,}

\title[Higher prevalence of X-ray selected AGN in intermediate age galaxies up to z$\sim$1]{AGN in intermediate age galaxies at z$\lesssim$1}

\author[A. Hern\'an-Caballero et al.]{
Antonio Hern\'an-Caballero,$^{1,2}$
Almudena Alonso-Herrero,$^{1,3}$
Pablo G. P\'erez-Gonz\'alez,$^{4,5}$\newauthor
Guillermo Barro,$^6$
James Aird,$^{7,8}$
Ignacio Ferreras,$^{9}$
Antonio Cava,$^{10}$ 
Nicol\'as Cardiel,$^4$\newauthor
Pilar Esquej,$^4$
Jes\'us Gallego,$^4$
Kirpal Nandra,$^{11}$
Javier Rodr\'iguez-Zaur\'in$^{12,13}$\\
$^{1}$Instituto de F\'isica de Cantabria, CSIC-UC, Avenida de los Castros s/n, 39005, Santander, Spain. E-mail: ahernan@ifca.unican.es\\
$^{2}$Augusto G. Linares Junior Research Fellow\\
$^{3}$Augusto G. Linares Senior Research Fellow\\
$^{4}$Departamento de Astrof\'isica y CC. de la Atm\'osfera, Facultad de CC. F\'isicas, Universidad Complutense de Madrid, E-28040 Madrid, Spain\\
$^{5}$Associate Astronomer at Steward Observatory, The University of Arizona, Tucson, AZ 85721, USA\\
$^{6}$UCO/Lick Observatory, Department of Astronomy and Astrophysics, University of California, Santa Cruz, CA 95064, USA\\
$^{7}$Department of Physics, Durham University, Durham DH1 3LE, UK\\
$^{8}$COFUND Junior Research Fellow\\
$^{9}$Mullard Space Science Laboratory, University College London, Holmbury St Mary, Dorking, Surrey RH5 6NT, UK\\
$^{10}$Observatoire de Gen\`eve, Universit\'e de Gen\`eve, 51 Ch. des Maillettes, 1290 Versoix, Switzerland\\
$^{11}$Department of Physics and Astronomy, Planck Institut f\"{u}r Extraterrestrische Physik, Giessenbachstra\ss e, 85748 Garching, Germany\\
$^{12}$Instituto de Astrof\'isica de Canarias, E-38200 La Laguna, Tenerife, Spain\\
$^{13}$Departamento de Astrof\'isica, Universidad de La Laguna, E-38205 La Laguna, Tenerife, Spain
}

\begin{document}
\date{Accepted ........ Received ........;}

\pagerange{\pageref{firstpage}--\pageref{lastpage}} \pubyear{2012}

\maketitle

\label{firstpage}

\begin{abstract}
We analyse the stellar populations in the host galaxies of 53 X-ray selected optically dull active galactic nuclei (AGN) at 0.34$<$z$<$1.07 with ultra-deep (m$_{AB}$$=$26.5, 3$\sigma$) optical medium-band (R$\sim$50) photometry from the Survey for High-z Absorption Red and Dead Sources (SHARDS).
The spectral resolution of SHARDS allows us to consistently measure the strength of the 4000 \AA{} break, \Dnu, a reliable age indicator for stellar populations.
We confirm that most X-ray selected moderate-luminosity AGN (L$_X$$<$10$^{44}$ erg s$^{-1}$) are hosted by massive galaxies (typically M$_*$ $>$10$^{10.5}$ M$_\odot$) and that the observed fraction of galaxies hosting an AGN increases with the stellar mass. A careful selection of random control samples of inactive galaxies allows us to remove the stellar mass and redshift dependences of the AGN fraction to explore trends with several stellar age indicators. We find no significant differences in the distribution of the rest-frame U-V colour for AGN hosts and inactive galaxies, in agreement with previous results. However, we find significantly shallower 4000\AA{} breaks in AGN hosts, indicative of younger stellar populations. 
With the help of a model-independent determination of the extinction, we obtain extinction-corrected U-V colours and light-weighted average stellar ages. We find that AGN hosts have younger stellar populations and higher extinction compared to inactive galaxies with the same stellar mass and at the same redshift. We find a highly significant excess of AGN hosts with \Dnu$\sim$1.4 and light weighted average stellar ages of 300--500 Myr, as well as a deficit of AGN in intrinsic red galaxies.
We interpret failure in recognizing these trends in previous studies as a consequence of the balancing effect in observed colours of the age-extinction degeneracy.

\end{abstract}


\section{Introduction} 

In the current paradigm of galaxy evolution, the growth of supermassive black holes (SMBH) and the galaxies that host them are intertwined \citep[see][, for a review]{Alexander12}. Observational evidence includes the tight correlation between the mass of the SMBH and the velocity dispersion in the bulge of the galaxies \citep[the so-called M-$\sigma$ relation;][]{Magorrian98, Ferrarese00,Gebhardt00,Marconi04} as well as a remarkable similarity between the redshift evolution of the cosmic star formation rate density and the integrated black hole accretion rate ($\dot M_{\rm BH}$), with both having their peak at $z$$\sim$1--3 and a steep decline from $z$$\sim$1 to the present \citep[e.g.,][]{Boyle98,Franceschini99,Merloni04,Chapman05,Merloni08,Silverman08,Bouwens09,Aird10}. In addition, active galactic nuclei (AGN), like star-forming galaxies, display a form of `downsizing' by which the bulk of SMBH growth and star formation shifts to lower luminosity or less massive systems at later epochs \citep{Cowie03,Fiore03,Hasinger05,Bongiorno07}.

Star formation takes place in one of two regimes. The majority of star formation \citep[up to 90\% at $z$$\sim$1--3;][]{Rodighiero11} occurs in secularly evolving systems, where internal processes (e.g. disc instabilities, turbulence) are responsible for gas dynamics that drive star formation \citep[e.g.][]{Elbaz07,Elbaz11,Tacconi08, Daddi10, Genzel10}. In these systems, the star formation rate (SFR) at a given redshift is roughly proportional to the galaxy mass, defining the so-called `main-sequence' \citep{Noeske07}.
However, a small fraction of galaxies sustain more efficient star formation in compact starbursts, which are commonly associated with mergers.  
Since both major mergers and internal processes are considered to be able to transport dust and gas to the inner regions of a galaxy \citep[e.g.][]{Kormendy04,Hopkins06}, finding the trigger for nuclear activity is not straightforward.
Early works suggested nuclear activity to be closely linked with major mergers, largely due to the high fraction of quasars that appear to be associated with merging systems
\citep[e.g.][]{Sanders88,Sanders96,Surace98,Canalizo01,Ivison10}.
However, later studies on the morphology of lower luminosity (L$_X$$<$10$^{44}$ erg s$^{-1}$) AGN hosts suggested that moderate levels of nuclear activity are typically associated with secular evolution rather than major mergers \citep[e.g.][]{Grogin05, Cisternas11, Schawinski11}.

The interplay between nuclear activity and star formation is not well understood. The luminosity and accretion rate of the most powerful AGN is found to correlate with the SFR in the host galaxy \citep[e.g.][]{Shi07,Chen13}, suggesting an important contribution of major mergers to the build up of the M-$\sigma$ relation. 
On the other hand, the majority of low and intermediate-luminosity AGN are not associated with major mergers, as many of them are hosted by ``normal'' discs \citep{Gabor09, Cisternas11, Ellison11, Schawinski11, Silverman11, Kocevski12}.

Albeit high resolution observations of local Seyferts have shown hints of a correlation between AGN activity and circumnuclear SFR in scales $\lesssim$1 kpc \citep[e.g.][]{Diamond-Stanic12,Esquej14}, several studies of moderate luminosity AGN at low and intermediate redshift find only a weak correlation with the SFR of the galaxy as a whole \citep{Silverman09,Shao10,Rosario12}. This is in qualitative agreement with results from simulations performed by \citet{Hopkins10}, which show an increasingly strong correlation of the SFR-$\dot M_{\rm BH}$ relation with decreasing physical scales from several kpc to $<$10 pc.
Further insight into the connection between AGN and star formation can be gained through the study of the stellar populations of the host galaxy. This usually requires to carefully remove the unresolved AGN component in ground-based images of local galaxies \citep[e.g.][]{Trump13} or, at higher redshifts, using HST data \citep{Jahnke04, Ammons11}. Another option is to select only low luminosity or obscured AGN that contribute a negligible fraction of the combined (AGN+galaxy) optical emission \citep[e.g.][]{Kauffmann03b,Alonso-Herrero08,Silverman09}.

Early studies showed that the rest-frame colours of AGN hosts are often in or close to the green valley, which led to speculation about the influence of the AGN in the transition from the blue cloud to the red sequence \citep[e.g.][]{Nandra07,Salim07,Bundy08,Silverman08,Georgakakis08,Schawinski09,Cimatti13}.
Later works, however, recognized the importance of stellar mass selection effects when comparing the colours of active and inactive galaxies. Some of these works found that AGN host colours are similar to those of inactive star-forming galaxies for the same mass and redshift \citep{Xue10,Rosario13}, while others suggested they are slightly redder \citep[e.g.:][]{Bongiorno12}.
Conflicting results have been associated at least in part to biases in the AGN or non-AGN control samples \citep{Xue10,Aird12,Rovilos12,Rosario13}. The strong evolution in the frequency of AGN detection with the stellar mass and redshift of the host, and AGN luminosity, implies that all three parameters need to be carefully controlled for meaningful comparisons between samples.

One basic difficulty in comparing the stellar populations of AGN hosts and inactive galaxies through rest-frame colours is that they depend not only on stellar age, but also on metallicity and extinction. This degeneracy implies that age differences can be either exaggerated or masked by differences in extinction. Extinction-corrected colours based on SED-fitting with libraries of synthetic templates can in principle solve this issue \citep[see e.g.][]{Cardamone10}, albeit at the cost of the results becoming model-dependent \citep[][, hereafter HC13]{Hernan-Caballero13}.

The strength of the 4000\AA{} break, \Dnu, and the H$_\delta$ line are two well known spectral indicators of stellar age \citep{Balogh99,Kauffmann03b}. Using a large sample of Sloan Digital Sky Survey (SDSS) spectra from local galaxies ($z$$<$0.3), \citet{Kauffmann03a} calibrated the star formation history (SFH) of low-z emission-line selected AGN. They found that the typical stellar ages of AGN hosts are younger than those of inactive galaxies while their mean SFR are higher.
Post-starburst spectroscopic signatures are also found to be strong in local AGN hosts \citep{Wild07}, and there is mounting evidence for the AGN activity peaking a few hundred Myr after the star formation does \citep{Davies07,Wild10,Alonso-Herrero13}.
At higher redshifts, \citet{Silverman09} demonstrated that \Dnu, the restframe U-V colour, and the SFR (based on the [OII] 3727\AA{} line) of a bright sample ($i_{acs}$$<$22.5) of X-ray selected AGN hosts are all consistent with each other, and match those of younger star-forming galaxies at the same redshift. However, spectroscopic surveys are limited to brighter magnitudes that do not include the bulk of the X-ray selected samples, which peak at fainter magnitudes.

Recently, several intermediate band optical and near infrared surveys have provided deep photometry with enough spectral resolution to infer the strength of the 4000 \AA{} spectral break at $z$$\lesssim$1 (HC13) and at higher redshifts \citep{Kriek11,Straatman14}.
In HC13 we analysed the stellar populations of a mass-selected sample of galaxies in the GOODS-N field with intermediate band photometry taken with the 10.4 m GTC telescope from the Survey for High-z Absorption Red and Dead Sources \citep[SHARDS;][]{Perez-Gonzalez13}. 
The SHARDS filterset consists of 24 contiguous medium-band (R$\sim$50) optical filters spanning the range 500--950 nm. SHARDS provides an uniform depth of $m$=26.5, (3$\sigma$) with sub-arcsec seeing in all its filters. 
We showed that measurements of the \Dn index on the SHARDS photospectra agree within $\sim$10\% with those obtained from full resolution spectra  (see Figure A4 in HC13), while they prove fainter magnitudes than the deepest spectroscopy available.
We also showed that, when combined with the rest-frame U-V colour, \Dn provides a powerful diagnostic of the extinction affecting the stellar population that is relatively insensitive to degeneracies with age, metallicity or star formation history. Using this novel approach, we estimated de-reddened colours and light-weighted stellar ages for individual sources. We explored the relationships linking stellar mass, rest-frame (U-V) colour, and \Dn for the non-active sources in the sample, and compared them to those found in local galaxies. 

In this work we study the stellar populations in the host galaxies of X-ray selected AGN in the redshift range 0.34$<$$z$$<$1.07 and within the fraction of the GOODS-N field covered by the SHARDS survey.
We compare rest-frame colours, \Dn indices, and average stellar ages of the AGN hosts with those of carefully matched comparison samples of inactive galaxies with the same underlying redshift and mass distributions.
The outline of the paper is as follows: \S\ref{sample-selection-sec} describes the selection of the AGN sample and the comparison sample of inactive galaxies. \S\ref{sect-analysis} deals with the obtention of stellar masses, rest-frame colours, \Dn indices, and average stellar ages. \S\ref{sect-results} presents our results regarding the stellar populations of AGN hosts. \S\ref{sect-discussion} discusses the systematics that could influence our results. \S\ref{sect-conclusions} summarizes our conclusions.
Throughout this paper we use a cosmology with $H_0$ = 70 km s$^{-1}$ Mpc$^{-1}$, $\Omega_M$ = 0.3, and $\Omega_\Lambda$ = 0.7. All magnitudes refer to the AB system.

\section{Sample selection}\label{sample-selection-sec}

Our parent sample is the catalog of X-ray sources in the 2 Ms Chandra Deep Field North (CDF-N) from \citet{Alexander03}. To minimize the number of contaminating sources that are not AGN, we constrain the sample to sources with a hard X-ray (2--8 keV) detection and located within the 141 arcmin$^2$ area of the GOODS-N field covered by SHARDS, which is centred very close to the aim point of the Chandra observations. There are 161 hard X-ray sources in the SHARDS area. The 3-$\sigma$ sensitivity limit of the Chandra observations in the 2--8 keV band is $\sim$10$^{-16}$ erg cm$^{-2}$ s$^{-1}$ at the centre of the SHARDS field and $\sim$3.3$\times$10$^{-16}$ erg cm$^{-2}$ s$^{-1}$ near the edges.

We find optical and IR counterparts to the Chandra sources using the likelihood ratio (LR) method \citep{Ciliegi03, Ciliegi05, Brusa07}. Very briefly, we consider all IRAC 3.6\um  sources within 3" of the Chandra position that are identified using the code developed for the Rainbow surveys \citep{Perez-Gonzalez08,Barro11a,Barro11b}. This code provides deblended IRAC photometry on the basis of higher resolution optical imaging \citep[see][ for further details]{Barro11a}. We calculate the LR for all candidate counterparts and determine a threshold in the LR that maximizes the sum of the completeness and reliability \citep[see][]{Luo10}. We deem any sources that exceed this LR threshold as secure counterparts (taking the counterpart with the highest LR value in cases of $>$1 secure candidate). We then repeat the entire method with candidates identified in the Subaru $R$, ACS $i$, and WIRCAM $K$ band images, retaining any additional secure counterparts identified in these bands. The Rainbow database\footnote{https://rainbowx.fis.ucm.es} compiles photometry (from UV to radio), redshifts, and SED-fitting derived physical parameters for galaxies in several cosmological fields, including GOODS-N. This ensures that counterparts are already cross-matched across all the optical/IR bands and provides consistent, matched-aperture photometry across all bands for our sources. This method will be described in more detail in Nandra et al. (in preparation). 96\% of all Chandra sources get a secure counterpart with this procedure.

Out of the parent sample of 161 hard X-ray sources, 115 have spectroscopic redshifts. For the remaining ones we rely on photometric redshifts obtained from the analysis of the broad-band and SHARDS SED (see \S\ref{subsect-masses-redshifts}).
Rest-frame 2--8 keV luminosities not corrected for absorption, L$_X$, are calculated following \citet{Trouille08} from the observed 2--8 keV flux densities, with K-corrections assuming a power-law spectrum with photon index $\Gamma$=1.8.

\begin{figure}
\begin{center}
\includegraphics[width=8.5cm]{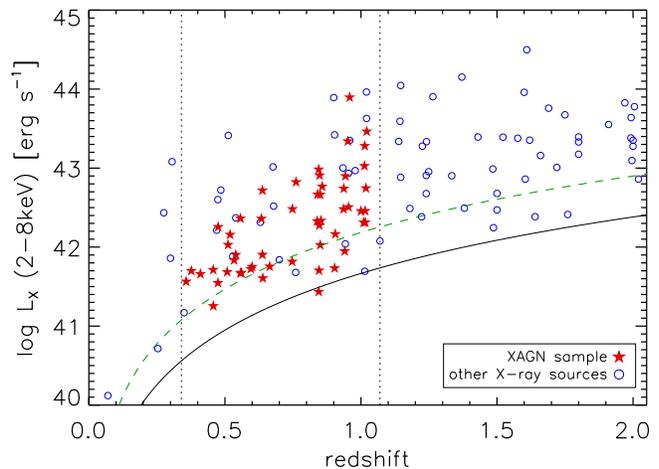}
\end{center}
\caption[]{Rest-frame hard X-ray (2--8 keV) luminosity (not corrected for absorption) versus redshift for the Chandra sources with detection in the 2--8 keV band in the SHARDS area. Stars represent the 53 sources selected for our stellar population analysis, while open circles mark other hard X-ray detected sources. The solid and dashed lines represent the 3-$\sigma$ sensitivity limits of the 2--8 keV Chandra maps at the centre and  edges of the SHARDS area, respectively. The dotted lines enclose the redshift range where the 4000 \AA{} break is observable in the SHARDS data.\label{LX-z-distr}}
\end{figure}

To ensure a reliable measurement of the 4000 \AA{} break in the SHARDS photospectra, we further constrain the sample to include only the galaxies with redshift 0.34$<$$z$$<$1.07. There are 75 such sources (68 of them with spectroscopic redshifts), with hard X-ray luminosities in the range $\log$(L$_X$/erg s$^{-1}$) $\sim$ 41.0--44.0 (Figure \ref{LX-z-distr}). 

Unobscured AGN can contribute a significant fraction of the rest-frame optical and NIR output of the galaxy. This causes dilution of the 4000 \AA{} break and 1.6 \um stellar bump, which may lead to underestimated stellar ages and overestimated stellar masses, respectively (see \S\ref{sect-discussion}).
To minimize the impact of AGN emission in the determination of stellar properties in the host galaxy, we have inspected Hubble Space Telescope (HST) Advanced Camera for Surveys (ACS) images of the 75 sources at 0.34$<$$z$$<$1.07 and removed from the sample 6 galaxies with bright point sources in their cores. However, powerful AGN can still dominate the NIR emission even if heavily obscured in the optical. \citet{Perez-Gonzalez08} concluded that the impact of the AGN on stellar mass estimates is significant only for X-ray sources with observed (i.e., not corrected for absorption) luminosities of L$_X$ $\gtrsim$ 10$^{44}$ erg s$^{-1}$ \citep[see also][]{Alonso-Herrero08}. Since our targets have lower L$_X$, stellar masses are expected to be accurate.
Even so, the X-ray to IR luminosity ratio varies significantly from one AGN to another \citep[see e.g.][]{Bongiorno12}, and the contribution of a moderate-luminosity AGN to the NIR emission could be substantial in the less massive hosts. 
Out of caution, we performed visual inspection of the infrared SED of the 75 X-ray selected sources in order to identify sources with red IRAC SEDs and no clear 1.6 \um bump. We found 11 such sources, 5 of them already discarded due to point-like cores in the ACS images.
In addition, there are 2 sources located very close to bright stars, and 7 more in the edges of the SHARDS images. These where removed from the sample due to incomplete SHARDS photometry. Finally, 1 source was removed because its spectroscopic redshift is inconsistent with the observed photometry and its photometric redshift is outside the 0.34$<$$z$$<$1.07 range.
The remaining 53 sources make up the final X-ray selected sample of optically faint AGN analysed in this work (hereafter, the XAGN sample). All the XAGN sources except 2 have spectroscopic redshifts. 

The hard X-ray emission in some of the faintest sources in the sample could originate from intense star formation instead of AGN activity. \citet{Lehmer12} estimate that at the flux limit of our sample (10$^{-16}$ erg s$^{-1}$ cm$^{-2}$) the fraction of inactive galaxies is $\sim$10\%.
We have calculated hard X-ray to optical luminosity ratios, L$_X$/L$_R$, and X-ray derived SFR, SFR$_X$, \citep{Vattakunnel12} for those sources with L$_X$$<$10$^{42}$ erg s$^{-1}$.
All but 6 sources show $\log$(L$_X$/L$_R$) $>$ -1 or SFR$_X$/SFR$_{UV+IR}$ $>$ 4 and are thus compatible with most of their hard X-ray emission arising from an AGN. Five of the 6 remaining sources have $f$(2-8keV)/$f$(0.5-2keV) between 2.5 and 8, consistent with a mildly to strongly absorbed AGN spectrum.
The remaining source has $f$(2-8keV)/$f$(0.5-2keV) $\sim$ 1, consistent with a starburst origin. In addition, it has the highest far-IR luminosity among the low luminosity X-ray sources, with SFR$_{UV+IR}$ = 126 M$_\odot$ yr$^{-1}$. 
However, the optical spectrum of this source shows [OIII] 5007$\AA$/H$_\beta$ $>$ 6. Accordingly, we believe this source hosts a moderate luminosity AGN, even if it does not dominate the X-ray emission.

\subsection{The reference sample of inactive galaxies}\label{reference-sample-subsec}

It is well known that the SFH of the galaxies is tightly linked to their stellar mass \citep[e.g.][]{Kauffmann03b,Xue10}, and that the fraction of AGN detections grows steeply as a function of the stellar mass of the host galaxy \citep{Kauffmann03b,Best05,Alonso-Herrero08,Bundy08,Silverman09,Mendez13}.
Therefore, to compare the stellar populations of AGN-hosts and inactive galaxies, it is of high importance to ensure that the mass distributions are the same.

We select a reference sample of inactive galaxies containing all the SHARDS sources with M$_*$$>$10$^9$M$_\odot$ and 0.34$<$$z$$<$1.07 that are not detected in X-rays. There are 2579 such sources. We note that the depth of the SHARDS data ensures a high mass completeness down to the mass limit of the sample ($>$97\% at $z$$\sim$1, see HC13). Since all the XAGN galaxies have stellar masses in the range 9.5$<$$\log$(M$_*$/M$_\odot$)$<$11.5 (see \S\ref{mass-dependence-subsec}), the reference sample contains virtually all the galaxies in the mass and redshift ranges probed by the XAGN sources. 

We build random comparison samples of inactive galaxies using a bootstrapping method similar to that described by \citet{Rosario13}. For each galaxy in the XAGN sample, we choose at random, and allowing repetitions, an inactive galaxy among those within $\pm$0.2 dex in stellar mass \citep[comparable to the typical uncertainty in stellar mass estimates from Rainbow,][]{Perez-Gonzalez08} and $\pm$0.1 in $z$. Since the reference sample is much larger compared to the XAGN sample, the number of repetitions is negligible.

We compare the distributions for any physical parameter in the XAGN galaxies and inactive galaxies by running Monte Carlo simulations in which 1000 random comparison samples are produced, all of them containing as many sources as the XAGN sample, with the same mass and redshift distributions. 
We obtain the distribution for the physical parameter in the population of inactive galaxies with the same underlying distributions of stellar mass and redshift of the XAGN galaxies as the mean of the distributions for the 1000 random samples. The dispersion of the distributions for individual random samples provides uncertainties for the mean distribution.  

To determine whether the distributions for the XAGN and inactive galaxies are compatible with being drawn from the same parent population, we run two-sample KS tests with the XAGN sample and each of the 1000 random comparison samples. The frequency distribution of the test $p$-value shows how often the null hypothesis (that the XAGN and inactive galaxies are drawn from the same parent population) can be rejected at a given confidence level.
If the null hypothesis is true, it should be rejected at the 0.05 confidence level in roughly 5\% of realizations. If the fraction of rejections is much higher it indicates the underlying distributions for the XAGN and inactive galaxies are different.
 
\section{Analysis}\label{sect-analysis}
\subsection{Stellar masses, redshifts, and rest-frame colours}\label{subsect-masses-redshifts}

\begin{figure*}
\begin{center}
\includegraphics[width=15cm]{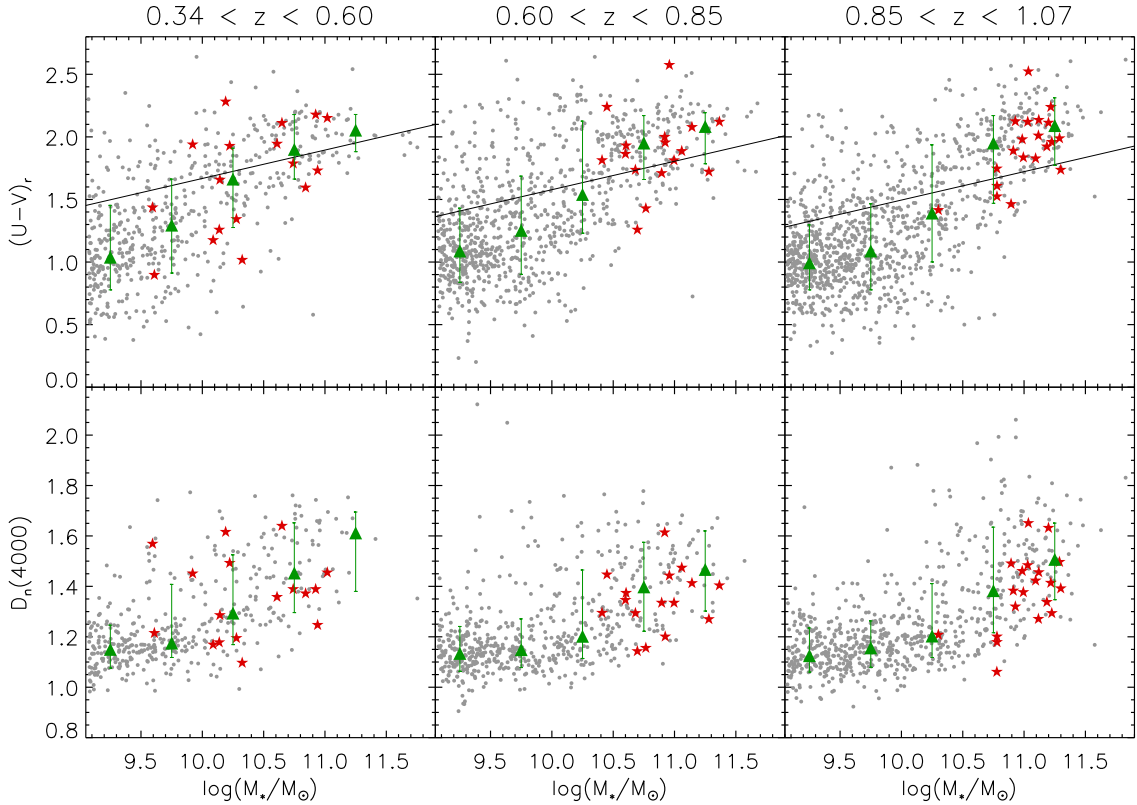}
\end{center}
\caption[]{dependence of the rest-frame (U-V) colour (top row), and \Dn index (bottom row) with the stellar mass for galaxies in three redshift bins. Small dots represent all the inactive galaxies in the reference sample (see \S\ref{reference-sample-subsec}), while stars represent sources in the XAGN sample. The green triangles with error bars represent median values and 16$^{th}$ and 84$^{th}$ percentiles in stellar mass bins of width 0.5 dex. The solid lines in the three upper panels represent the location of the green valley at the redshift corresponding to the centre of each interval.\label{colorsmassx9}}
\end{figure*}

We obtain stellar masses, redshifts, and rest-frame colours from the Rainbow database.
The procedure used to estimate photometric redshifts and stellar masses is similar to the one described in \citet{Perez-Gonzalez08}, with a few adaptations to include the intermediate band photometry from SHARDS (P\'erez-Gonz\'alez et al., in preparation). Very briefly, we use a maximum likelihood estimator to find the stellar population synthesis (SPS) model that best fits all the available photometric data points for wavelengths $<$4\um (rest-frame). The stellar emission in the models is taken from the PEGASE code \citep{Fioc97} assuming a \citet{Salpeter55} initial mass function (IMF) with 0.1$<$M$_*$/M$_\odot$$<$100, and a SFH described by a declining exponential law with timescale $\tau$ and age $t$ [i.e, SFR($t$) $\propto$ $e^{-t/\tau}$]. The uncertainty in the stellar mass of individual galaxies is $\sim$0.2--0.3 dex, while the dispersion in the photometric redshifts is $\Delta$($z$)/(1+$z$)$\sim$0.0067. 

Rest-frame magnitudes in several UV and optical broad-band filters are computed by convolving the best-fitting SPS model with the filter transmission curve, as described in \citet{Perez-Gonzalez08}. The broad spectral coverage of the Rainbow photometry implies that synthetic photometry in the rest-frame $U$, $V$, and $J$ bands is interpolated between observed filters. Owing to the accurate photometric redshifts, the uncertainty in rest-frame colours is comparable to the uncertainty in observed colours, $\sim$0.1 magnitudes at $m$=25.5.

The first row in Figure \ref{colorsmassx9} shows the dependence of the restframe (U-V) colour, (U-V)$_r$ with the stellar mass for three redshift bins. The solid lines represent the location of the green valley at the middle of each redshift interval, as derived from the relation obtained by \citet{Borch06} in a large sample of 0.2$<$$z$$<$1.0 galaxies from the COMBO-17 survey \citep{Wolf03}.
The green valley is defined by:
\begin{equation}
(U-V)_{GV} = 0.227 \log(M_*/M_\odot) - 0.352 z - 0.437
\end{equation}
\noindent where the value of the constant term has been adjusted for magnitudes expressed in the AB system and stellar masses derived assuming a \citet{Salpeter55} IMF.

\subsection{\Dn index, extinction-corrected colours, and average stellar ages}

We measure the strength of the 4000 \AA{} break, \Dnu, in the SHARDS photospectra using the definition of \citet{Balogh99}. Raw \Dn values are corrected for the limited spectral resolution of the SHARDS photospectra (R$\sim$50) using the calibration obtained in HC13 for synthetic SHARDS photometry extracted from high resolution spectra. The typical uncertainty in \Dn estimates is $\sim$10\%. 

The bottom row in Figure \ref{colorsmassx9} shows the \Dn index versus the stellar mass for galaxies in three redshift bins. Compared to the restframe U-V colour, \Dn is much less sensitive to the stellar mass for M$_*$$<$10$^{10.5}$M$_\odot$. Blue cloud galaxies show higher dispersion in (U-V)$_r$ compared to \Dnu, arguably as a consequence of the higher impact that extinction has on broadband colours (see HC13 for a discussion).
 
We combine information from (U-V)$_r$, and \Dn to compute extinction-corrected (intrinsic) values, (U-V)$_0$ and \Dno. We use the method of descent down to the dust-free sequence described in HC13. The method relies on the tight correlation between (U-V)$_0$ and \Dno{} observed in model SEDs and the universality of the ratio $E$($U$-$V$)/$\Delta \log$\Dn irrespective of the extinction law.
Unlike SED-fitting based extinction corrections, this method is largely independent of assumptions about the metallicity or SFH of the galaxy, and its results are not affected by the usual degeneracy among age, extinction, and metallicity. See \S4.5 in HC13 for further details. Uncertainties in (U-V)$_0$ range from 0.2 to 0.4 magnitudes, with the typical value being 0.3 magnitudes.

\begin{figure}
\begin{center}
\includegraphics[width=8.5cm]{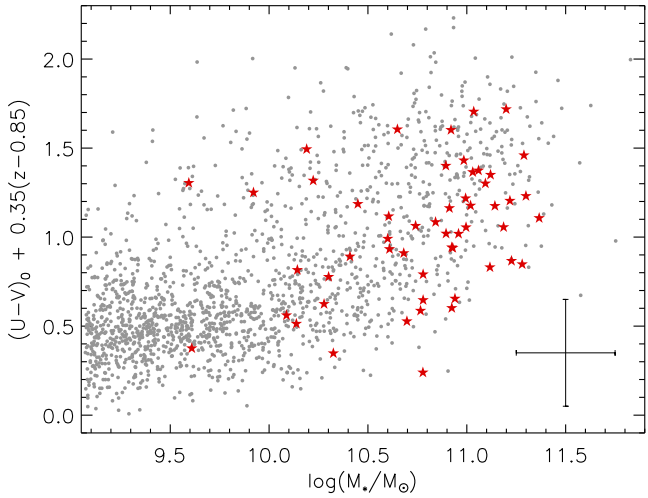}
\end{center}
\caption[]{Restframe (U-V) colour corrected for extinction and for the redshift evolution of the red sequence, versus the stellar mass. Symbols as in Figure \ref{colorsmassx9}. The error bars in the bottom-right corner represent typical 1-$\sigma$ uncertainties for individual sources.\label{UV0z-smass}}
\end{figure}

Figure \ref{UV0z-smass} represents the extinction-corrected restframe (U-V) colour, (U-V)$_0$, with an additional correction term $\Delta$(U-V) = 0.352($z$-0.85) that accounts for the redshift evolution of the red sequence. The less massive galaxies (M$_*$$<$10$^{10}$ M$_\odot$) concentrate in a narrow blue cloud around (U-V)$_0$ + $\Delta$(U-V) $\sim$ 0.5, while more massive galaxies show an increasingly red colour.
We observe no strong bimodality in the distribution of (U-V)$_0$, in contrast to results from \citet{Cardamone10} in a similar sample.
While the relatively large uncertainties in (U-V)$_0$ could in principle smooth the bimodality by filling with sources a narrow green valley, the $\sim$1 magnitude difference that \citet{Cardamone10} find between the extinction-corrected colour of red sequence and blue cloud galaxies would easily show up in our data (see \S4.5 in HC13 for a discussion).
The XAGN sources have (U-V)$_0$ values within the same range of inactive galaxies at the same stellar mass, however, they seem to concentrate at intermediate values, particularly in the most massive galaxies. For a detailed, quantitative analysis see \S\ref{stellar-ages-subsec}.

We define the light-weighted average stellar age, $t_{ssp}$, for individual galaxies as the age of the single stellar population (SSP) model that produces the same extinction-corrected \Dnu. We used the SSP model library of \citet{Bruzual03} with a \citet{Salpeter55} IMF and solar metallicity. We emphasize that $t_{ssp}$ is not supposed to represent the formation age of the galaxy or the age of any particular stellar population within it. Instead, it is a convenient way to represent extinction-corrected \Dn measurements calibrated in units of time. See HC13 for further details on how $t_{ssp}$ relates to the SFH. 

\section{Results}\label{sect-results} 

\subsection{Mass dependence of the AGN fraction}\label{mass-dependence-subsec}

Multiple studies show a steep increase in the fraction of AGN detections with the stellar mass of the host. This trend is observed independently of the method employed to select the AGN: X-ray selection \citep{Alonso-Herrero08,Bundy08,Silverman09,Mendez13}, optical emission lines \citep{Kauffmann03b}, or radio-loudness \citep{Best05}.
The steep increase in the AGN fraction with the stellar mass has been considered a selection effect due to the M-$\sigma$ relation, which implies that at a given Eddington ratio more massive galaxies host more luminous AGN. The shape of the underlying distribution of Eddington ratios is however thought to be independent of the stellar mass and redshift \citep[e.g.,][]{Aird12,Bongiorno12}. 

\begin{figure}
\begin{center}
\includegraphics[width=8.5cm]{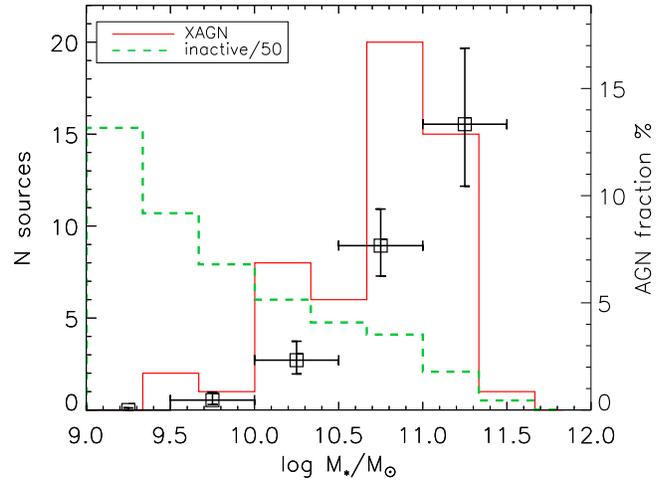}
\end{center}
\caption[]{Distribution of stellar masses for the X-ray selected optically faint AGN (solid line) and the inactive galaxies (dashed line). The number counts for inactive galaxies have been scaled down by a factor 50 to improve readability. Open symbols represent the AGN fraction in bins of stellar mass 0.5 dex wide with the scale indicated in the right-hand side of the graph. Horizontal bars indicate the mass range of each bin, while vertical bars represent the 68\% confidence interval of the AGN fraction calculated with the Wilson formula for binomial distributions.\label{mass-Silverman}}
\end{figure}

In the SHARDS sample we also find that X-ray selected AGN are preferentially hosted by massive galaxies, with 50\% of the AGN hosts at $>$10$^{11}$ M$_\odot$ and $\sim$95\% over 10$^{10}$ M$_\odot$.
Figure \ref{mass-Silverman} compares the stellar mass distribution of the XAGN sources and the reference sample of inactive galaxies. The fraction of galaxies that host an X-ray selected AGN is less than 1\% for galaxies below 10$^{10}$ M$_\odot$, but increases to $\sim$13\% for galaxies over 10$^{11}$ M$_\odot$.

The frequency of AGN as a function of stellar mass has been discussed by \citet{Silverman09}. They selected a sample of X-ray sources using similar criteria in the much wider but shallower XMMCOSMOS survey (50 ks compared to 2 Ms in the CDF-N). Their sample is limited to galaxies brighter than $i$=22.5 AB (the faint limit of their spectroscopic sample) and $\log$(M$_*$/M$_\odot$)$>$10.6. In addition, they required for selection X-ray fluxes above the thresholds 5$\times$10$^{-16}$ or 2$\times$10$^{-15}$ erg s$^{-1}$ cm$^{-2}$ in the 0.5--2 or 2--10 keV bands, respectively. In the redshift range 0.5$<$$z$$<$1.0 (sample B in their Table 1) the fraction of sources with X-ray luminosity L$_X$[0.5-10keV]$>$10$^{42}$ erg s$^{-1}$ is 4\%. If we restrict our sample to galaxies in the same redshift range and with 2--8 keV fluxes above 2$\times$10$^{-15}$ erg s$^{-1}$ cm$^{-2}$ we select only 10 out of 294 massive galaxies, or 3.4\%, in good agreement with the results of \citet{Silverman09}. However, the much higher depth of CDF-N observations (3$\sigma$ detection limit of $\sim$10$^{-16}$ erg s$^{-1}$ cm$^{-2}$ in the centre of the field) increases our AGN fraction to 12\% of the $\log$(M$_*$/M$_\odot$)$>$10.6 galaxies when the fainter X-ray sources are considered. 

\subsection{Rest-frame colours of AGN hosts}\label{restframe-colours-subsec}
 
\begin{figure*}
\begin{center}
\includegraphics[width=17.0cm]{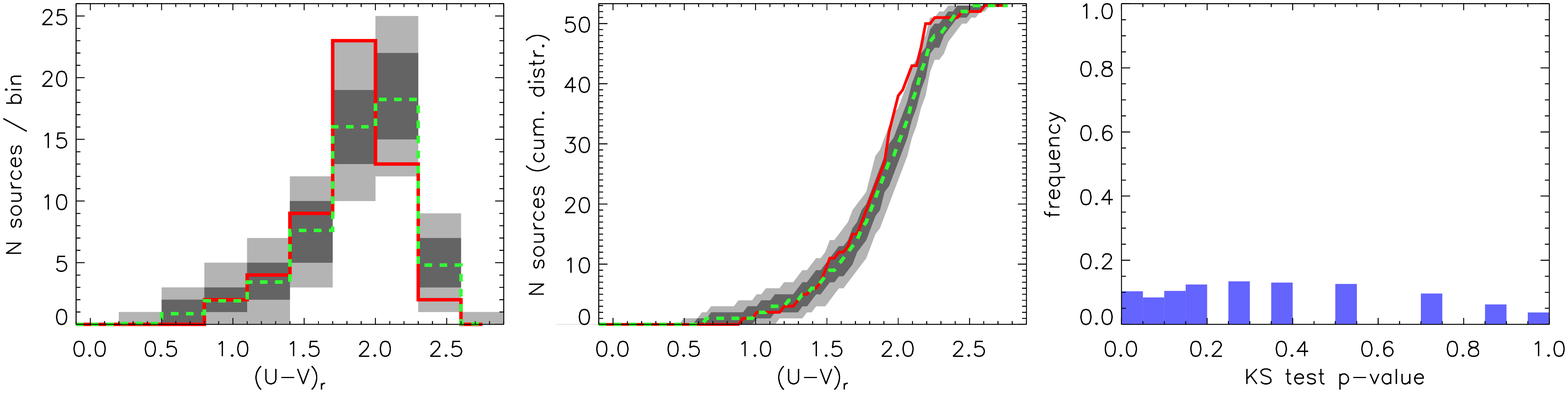}
\includegraphics[width=17.0cm]{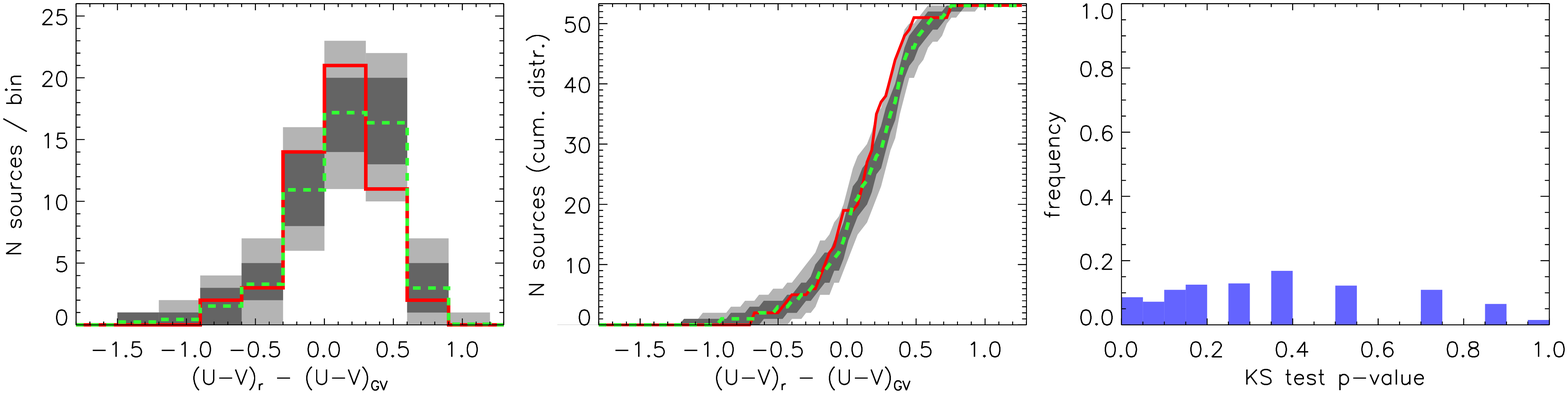}
\includegraphics[width=17.0cm]{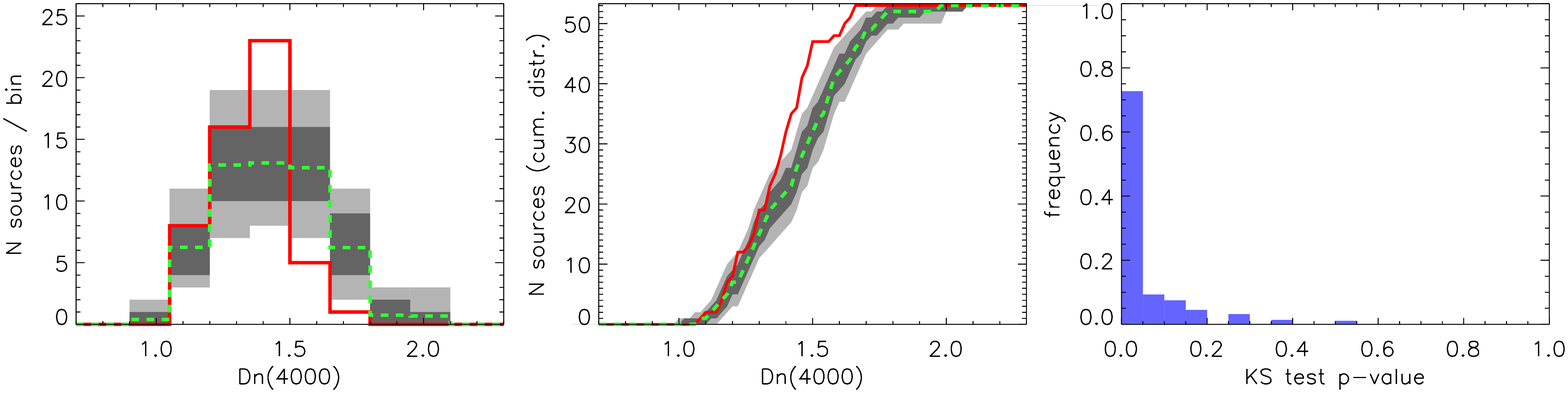}
\end{center}
\caption[]{Distributions for several stellar age indicators in the XAGN sample and the control samples of inactive galaxies. The parameters shown are: rest-frame U-V colour (top row), distance to the green valley (central row), and strength of the 4000 \AA{} break (bottom row). In each row the left panel shows number counts in equally sized bins, while the central panel shows cumulative distributions. The solid lines represent distributions for the XAGN sources, while the dashed lines represent the average of the 1000 random control samples. Dark and light grey areas represent the 1-$\sigma$ and 2-$\sigma$ dispersion of individual control samples, respectively. Histograms in the right panel show the frequency distribution of the KS-test p-values (see text for details).\label{AGNstats}}
\end{figure*}
 
We address first the question of whether the U-V colour of AGN hosts and inactive galaxies are different. 

The distribution of the rest-frame (U-V) colour for the AGN sample and the average of 1000 realizations of the control sample is shown in the top row of Figure \ref{AGNstats}. 
Both distributions show comparable shapes, with a peak in the red sequence at (U-V)$_r$$\sim$2 (albeit the distribution for the XAGN peaks at slightly bluer (U-V)$_r$) and an exponential tail towards the blue cloud.
The cumulative distribution for the XAGN is contained within the 2-$\sigma$ dispersion of the mean distribution for inactive galaxies, and the KS-tests finds the two distributions are not significantly different.

While our sample spans a large redshift range, the use of comparison samples matched in stellar mass and redshift ensures that a potential difference in the colour distribution of XAGN and inactive galaxies is not obfuscated by redshift and luminosity evolution. To prove this point, we calculate for every galaxy its distance to the green valley d$_{GV}$ = (U-V)$_r$ - (U-V)$_{GV}$, with the rest-frame colour of the green valley at the corresponding stellar mass and redshift, (U-V)$_{GV}$, calculated from Eq. 1.
The second row in Figure \ref{AGNstats} shows the distribution of distances to the green valley for the XAGN and inactive samples. Their shapes are very similar to the distributions for (U-V)$_r$, because (U-V)$_{GV}$ varies by only 0.2 magnitudes for a 10-fold increase in the stellar mass, and the redshift evolution is also small. 
About 2/3 of the sources in the XAGN and the comparison samples of inactive galaxies are in the red sequence (d$_{GV}$$>$0). The mean d$_{GV}$ for red sequence galaxies is slightly lower for the XAGN compared to the inactive galaxies (0.27 vs 0.32 magnitudes). 
However, this difference is not significant, and the KS-test finds the two distributions are compatible with being drawn from the same parent population.
 This is in agreement with recent results obtained in similar samples \citep[e.g.:][]{Xue10,Rosario13}, which found no significant differences between the rest-frame U-V colour of AGN hosts and inactive galaxies.

The similarity of the observed U-V colours in mass and redshift matched samples is striking given the widespread observation of a low prevalence of AGN among quiescent galaxies \citep[e.g.:][]{Kauffmann03b,Xue10,Schawinski11} and higher average SFR in AGN hosts compared to inactive galaxies \citep{Santini12,Rosario13}. These facts can be reconciled if AGN are more likely to be hosted in star-forming galaxies with significant extinction \citep[see e.g.][]{Cimatti13}. Since extinction reddens the U-V colour, it can push some dusty star-forming galaxies into the red sequence. These galaxies would show U-V colours similar to those of quiescent galaxies, but higher star formation rates.

\begin{figure*}
\includegraphics[height=7.1cm]{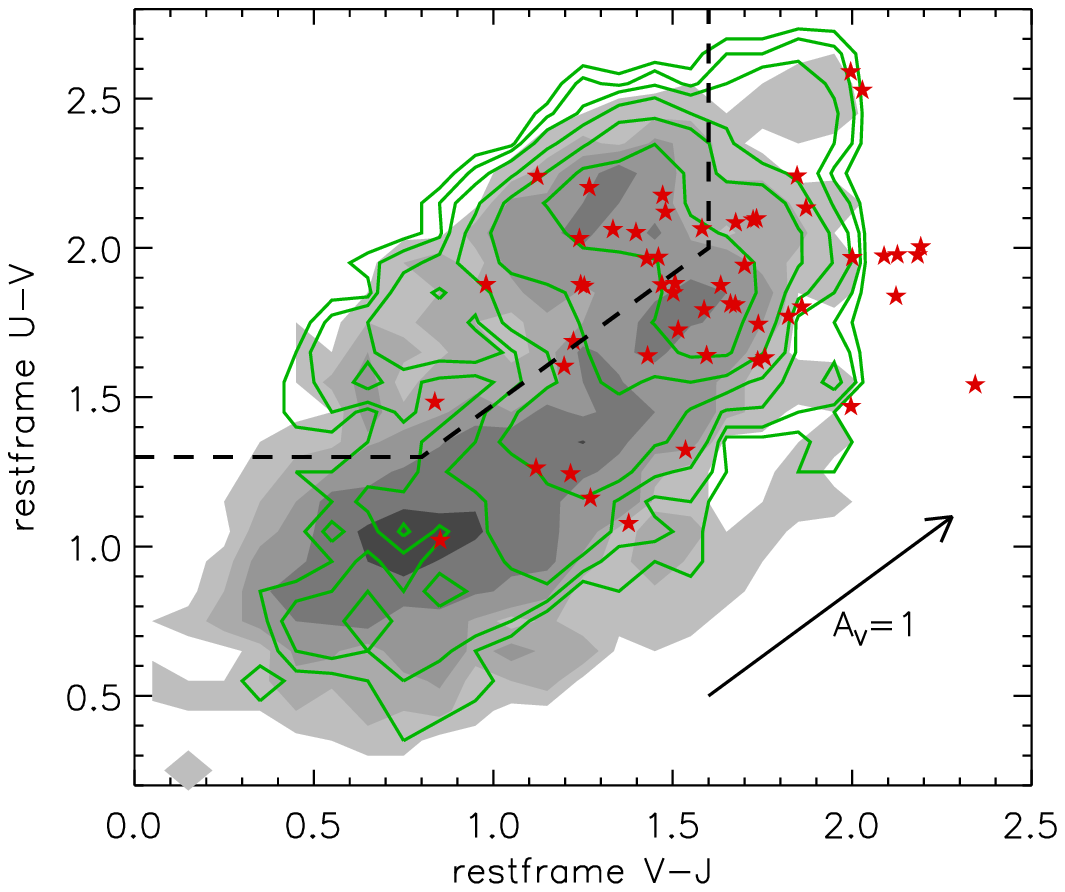}\hfill
\includegraphics[height=7.1cm]{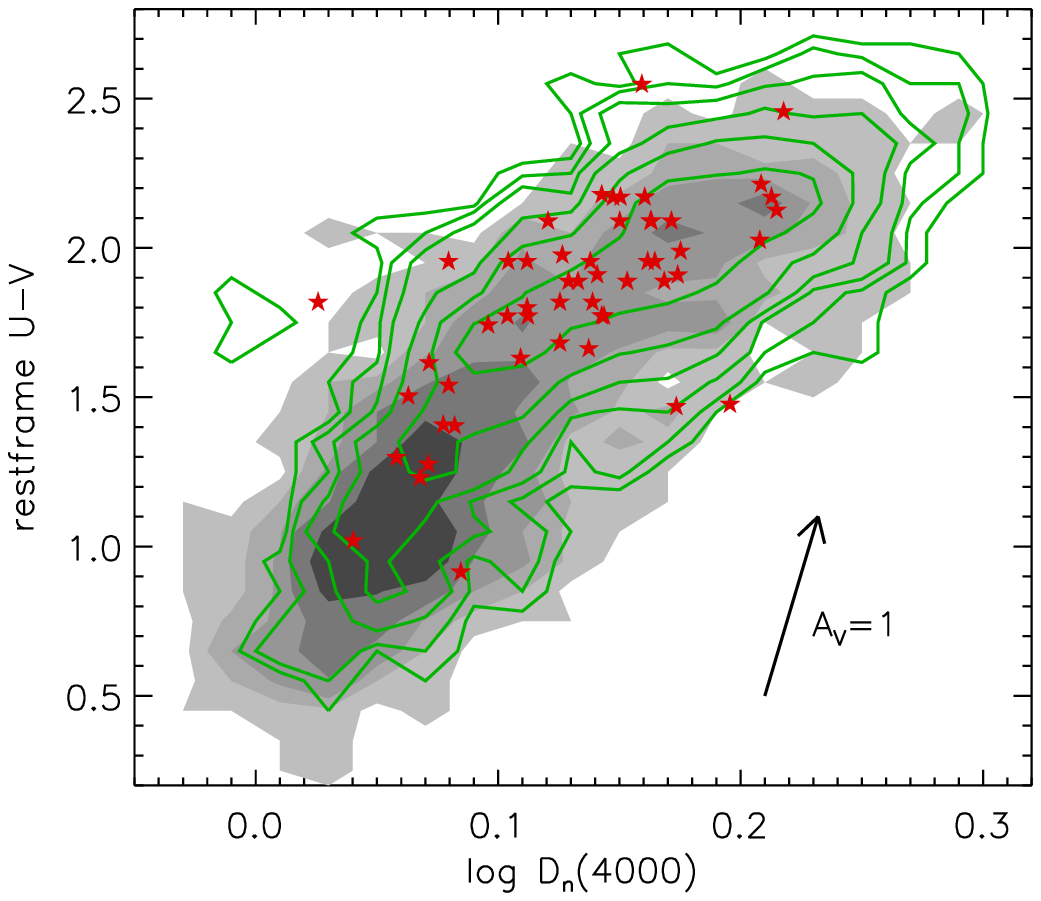}
\caption[]{UVJ rest-frame color-color diagram (left) and rest-frame U-V (not corrected for extinction) vs $\log$\Dn diagram (right). Stars represent sources in the XAGN sample, while 
the greyscale map represents the density distribution of the 2579 inactive galaxies in the parent comparison sample.
The average density distribution for the 1000 random samples matched in mass and redshift to the XAGN sample is shown as contours. The scale for isodensity lines is logarithmic, with each level representing twice the density of the previous one. The `locus' of quiescent galaxies is delimited by the dashed line in the UVJ diagram. The arrow in the bottom right corner indicates the effect in the rest-frame colours of 1 magnitude extinction in the $V$ band assuming a \citet{Draine03} extinction law with Milky Way grain size distribution \citep{Weingartner01} and R$_V$=3.1.\label{UVJ-plot}}
\end{figure*}

Evidence favouring higher extinction in the AGN hosts comes from the UVJ restframe color-color diagram (Figure \ref{UVJ-plot}, left panel). In this diagram, quiescent galaxies occupy the region with red (U-V)$_r$ and blue (V-J)$_r$ delimited by the dashed line \citep{Williams09}.
At the same (U-V)$_r$, reddened star-forming galaxies are about 0.7 magnitudes redder in the (V-J)$_r$ colour compared to quiescent galaxies.
Only 21\% of the XAGN sources are found in the locus of the quiescent galaxies, while the same region contains 44\% of the inactive galaxies in the random comparison samples. In addition, sources with (V-J)$_r$$>$1.7 represent 32\% of the XAGN compared to only 17\% of inactive galaxies. 

The 21\% fraction of passively evolving AGN hosts is much lower than the $\sim$50\% found by \citet{Cardamone10} at $z$$\sim$1 in the Extended Chandra Deep Field South, but consistent with recent results from \citet{Georgakakis14}, who find in a larger sample including the CDF-S that $\sim$15-20\% of the AGN luminosity density at z$\approx$0.40 and 0.85 is associated with galaxies in the quiescent part of the UVJ diagram.

\subsection{Stellar ages of AGN hosts}\label{stellar-ages-subsec}

\begin{figure*}
\begin{center}
\includegraphics[width=17.0cm]{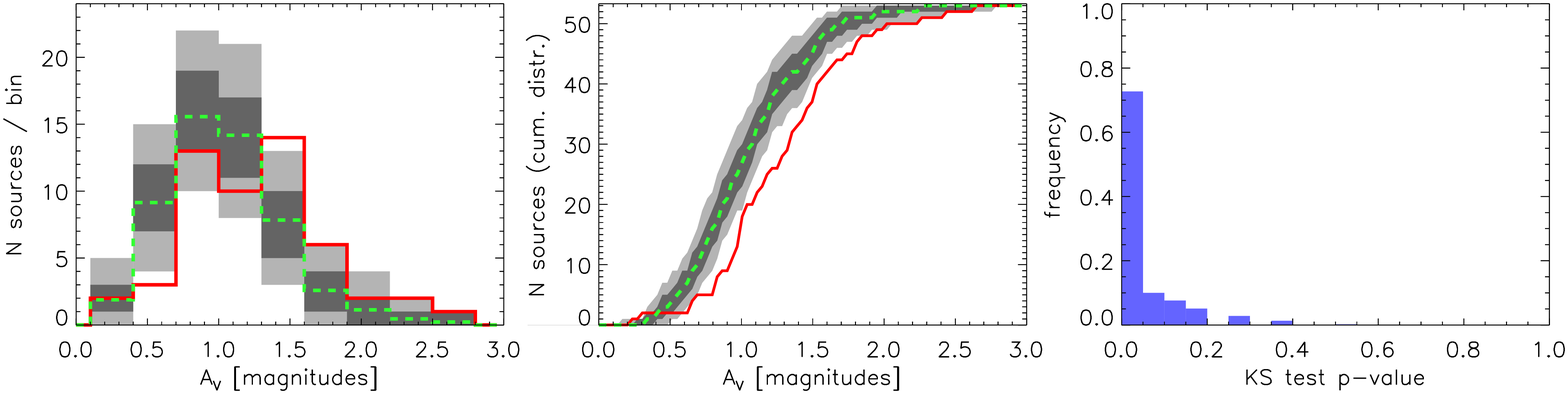}
\vspace{0.1cm}
\includegraphics[width=17.0cm]{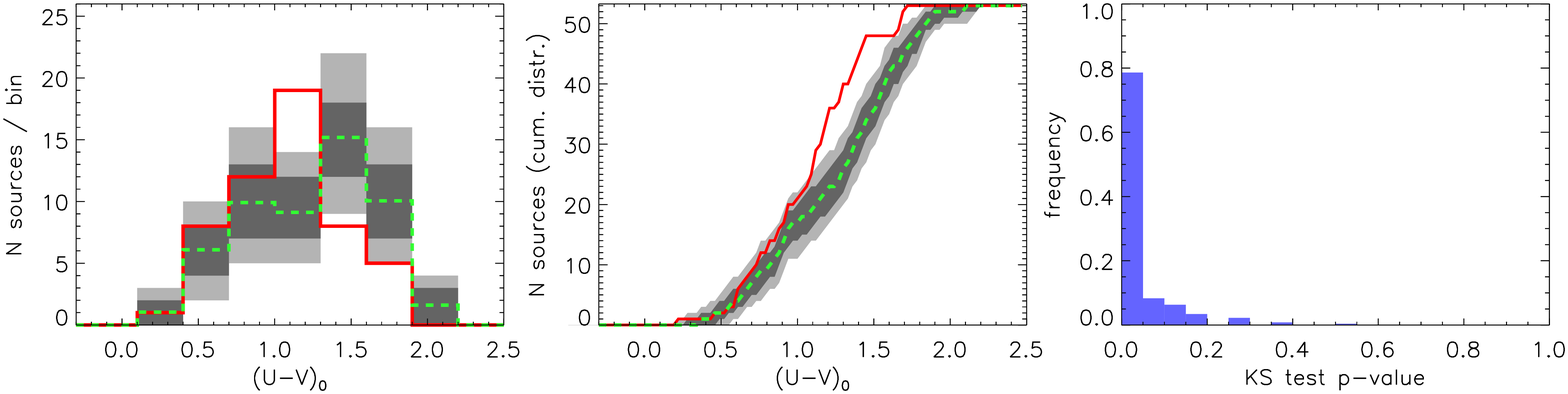}
\vspace{0.1cm}
\includegraphics[width=17.0cm]{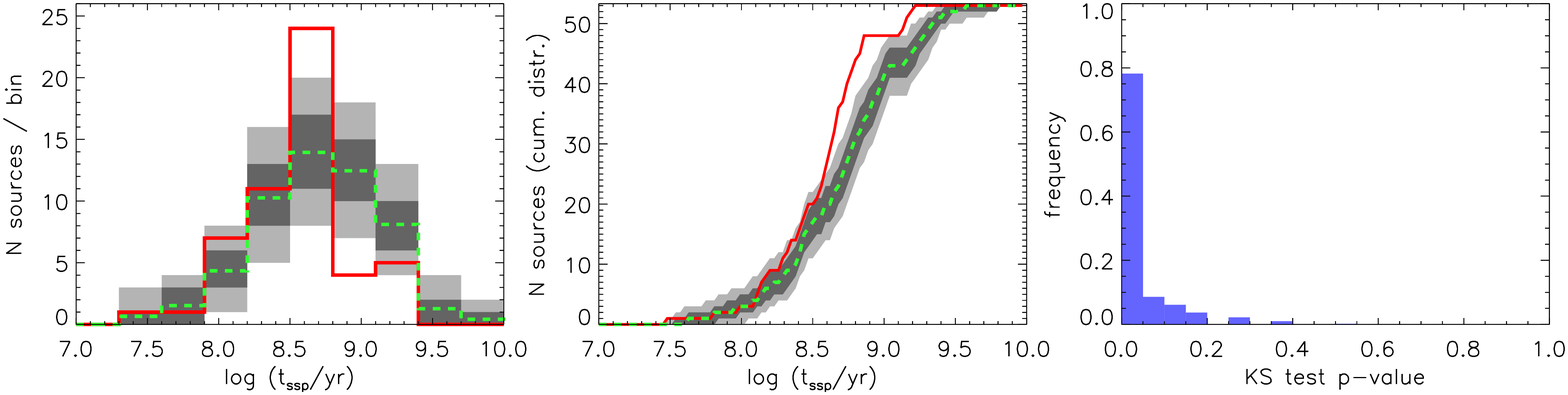}
\end{center}
\caption[]{Distributions of restframe $V$-band extinction estimates (top row), extinction corrected rest-frame (U-V) colour (central row), and light weighted average stellar age (bottom row). Symbols as in Figure \ref{AGNstats}.\label{AGNstats2}}
\end{figure*}

Unlike the (U-V)$_r$ colour, the \Dn index is only weakly influenced by extinction, and therefore offers a more robust indication of the age of stellar populations. 
The bottom row in Figure \ref{AGNstats} shows the distributions of \Dn for the XAGN and the control samples. The AGN counts are now outside the 2-$\sigma$ confidence interval for inactive galaxies in several \Dn bins. There is a clear ($>$3$\sigma$) excess of AGN in galaxies with \Dnu$\sim$1.4, and also a strong deficit of AGN at \Dnu$>$1.5. The KS-test confirms that the XAGN and control samples are different at the $\alpha$=0.05 significance level in 70\% of realizations. This is incompatible with the null hypothesis of the XAGN and control samples originating from the same population. The small fraction of XAGN sources with \Dnu$>$1.5 indicates a low prevalence of AGN among passively evolving galaxies. 
We note that the peak in the AGN number counts at \Dnu$\sim$1.4 is also apparent in M$_*$$>$10$^{10.6}$M$_\odot$ galaxies at comparable redshifts from zCOSMOS \citep[][; their Figure 11]{Silverman09}, although it is more prominent in our data.  

The (U-V)$_r$ versus $\log$ \Dn diagram (Figure \ref{UVJ-plot}, right panel) shows that there is a tight correlation between (U-V)$_r$ and \Dn that applies to both XAGN and inactive galaxies. However, about 1/3 of XAGN with intermediate \Dn values and (U-V)$_r$$\sim$2 show (U-V)$_r$ excesses of 0.2--0.4 magnitudes relative to inactive galaxies with the same \Dnu. Also, very few of the XAGN with (U-V)$_r$$\sim$2 have $\log$\Dnu$\sim$0.2, as is the case for the inactive galaxies. Instead, they have lower $\log$ \Dn values which imply the red (U-V)$_r$ is due to higher extinction. Since (V-J)$_r$ and \Dn are obtained with independent techniques using different data sets (broad-band versus intermediate band photometry), we are confident that the observed trends are real.

It is enlightening to test whether extinction estimates and extinction-corrected U-V colours have different distributions for the AGN hosts and inactive galaxies that support a lower prevalence of AGN among intrinsic red galaxies.
One difficulty with SED-fitting based extinction corrections is the degeneracy between metallicity, extinction, and stellar age. While this degeneracy can be partly broken with the help of rest-frame NIR photometry, this implies that the extinction determination is affected by the entire UV to NIR SED of the galaxy, not just the population that dominates the emission in the rest-frame $U$ and $V$ bands. 
To overcome this issue we obtain an extinction correction for (U-V)$_r$ and \Dn using the method of projection down to the dust-free sequence presented in HC13. This method has the advantages of relying only on the SED between the rest-frame $U$ and $V$ bands and not depending on assumptions about the metallicity or SFH of the galaxies.
The first row in Figure \ref{AGNstats2} shows the distribution of extinction in the rest-frame $V$-band calculated with this method for XAGN and inactive galaxies. While the uncertainties in A$_V$ are large ($\sim$0.3--0.4 magnitudes) there is a clear trend towards higher extinction in the XAGN, with a median value of 1.25 compared to 1.0 in the inactive galaxies.

The second row in Figure \ref{AGNstats2} shows the distribution of the extinction-corrected rest-frame U-V colour, (U-V)$_0$. There is a strong excess of AGN at (U-V)$_0$$\sim$1.2, whose significance ($>$4$\sigma$) is boosted by a slight decrease in the number counts of inactive galaxies at the same (U-V)$_0$. 
There is also a significant deficit of intrinsic red galaxies (extinction-corrected (U-V)$_0$$>$1.3) among the XAGN. The cumulative distributions for \Dn and the extinction-corrected (U-V)$_0$ have similar shapes. This is hardly surprising, since the two magnitudes are not independent. However, the significance of the difference between the XAGN and inactive galaxy distributions is even higher for the extinction-corrected colour (80\% of realizations with p-value$<$0.05), probably due to the residual effect that extinction has on \Dn values.

While the correspondence between the extinction-corrected values of (U-V)$_r$ and \Dn is independent of metallicity or SFH, to translate any of them to stellar ages requires to assume a metallicity and SFH. For simplicity, we define the light weighted average age of the stellar population, $t_{ssp}$, as the age of an instantaneous burst with solar metallicity that produces the observed extinction-corrected \Dno{} and (U-V)$_0$.
The bottom row in Figure \ref{AGNstats2} represents the distributions of $t_{ssp}$ for the XAGN and inactive galaxies. Since there is a functional relation between the extinction-corrected (U-V)$_0$ and $t_{ssp}$, it provides no new information. The small differences in the number counts, cumulative distributions and frequency distribution of p-values are due to differences in binning, the non-linear nature of the colour-$t_{ssp}$ relationship and the randomness of comparison samples.   

The \Dno{} and (U-V)$_0$ of the peak of the XAGN distribution translates to $t_{ssp}$$\sim$300--500 Myr. In HC13 we estimated from stellar population models that a galaxy with a constant SFR has a light-weighted average stellar age that converges towards $t_{ssp}$ = 300 Myr. This implies that galaxies with $t_{ssp}$$<$300 Myr must have experienced a recent increase in their SFR, while those with $t_{ssp}$$>$300 Myr have recent SFR below their long-term average. The observed peak in the frequency of X-ray detected AGN at 300$<$$t_{ssp}$$<$500 Myr would then imply that the probability of observing AGN activity peaks when the last star formation episode is already in decline. This is in agreement with the detection of significant post-starburst stellar populations in the hosts galaxies of luminous local AGN \citep{Kauffmann03b} and results from detailed analysis of the stellar populations in local AGN hosts, which find that the average accretion rate rises steeply $\sim$250 Myr after the onset of the starburst \citep{Wild10}.

In summary, the distribution of restframe (U-V) colours and distances to the green valley of XAGN sources are not significantly different from those of inactive galaxies, in agreement with previous results. However, once the effects of extinction are removed (either by using the \Dn index or the extinction corrected (U-V) colour) there is a clear excess of AGN at values that correspond to intermediate stellar ages ($t_{ssp}$=300--500 Myr) as well as a deficit of AGN in quiescent galaxies. 

\subsection{Mass dependence of the age distribution of AGN frequencies}

We have shown evidence that the frequency of AGN detections depends mainly on the stellar mass of the host galaxy. Once the stellar mass selection effects are taken into account, a clear dependence on stellar age appears that favours intermediate age hosts. We wish to address now the question of whether there is a mass dependence on the distribution of the AGN fraction as a function of stellar age.
Other studies have found evidence for such a dependence. In a larger sample of X-ray selected AGN, \citet{Aird12} found that for the most massive galaxies ($>$10$^{11}$ M$_\odot$) the AGN fraction is highest in the blue cloud, while at lower masses it peaks in the green valley. Using the \Dn index, \citet{Silverman09} obtained an equivalent result: while at M$_*$$>$10$^{11.1}$ M$_\odot$ the AGN fraction is higher among the galaxies with a weaker 4000 \AA{} break, the peak shifts to \Dn$\sim$1.5 if galaxies in the 10$^{10.6}$$<$M$_*$/M$_\odot$$<$10$^{11.1}$ M$_\odot$ range are also included.

For comparison, we show in Table \ref{XAGN-freq-table} the AGN fraction as a function of (U-V)$_r$, \Dnu, (U-V)$_0$, and $t_{ssp}$ for three mass bins. Percentages indicate the AGN fraction among galaxies within a given interval of stellar mass and one of the age indicators. The values in parenthesis represent the number of XAGN sources and the total number of galaxies in each group.

We find that for the mass bins 10.5$<\log$M$_*$/M$_\odot<$11.0 and 11.0$<\log$M$_*$/M$_\odot<$11.5 all four indicators are consistent with comparable AGN fractions for young and intermediate age stellar populations, and much lower fractions (a factor 3--5 lower) in galaxies with old populations. Although the AGN fraction seems to increase slightly faster with stellar mass for the galaxies with the youngest stellar populations, we find the difference not to be significant due to the limited sample size. Therefore, we cannot confirm a mass dependence of the relative frequency of AGN as a function of stellar age.

\subsection{dependence with X-ray luminosity}

\begin{figure}
\begin{center}
\includegraphics[width=8.5cm]{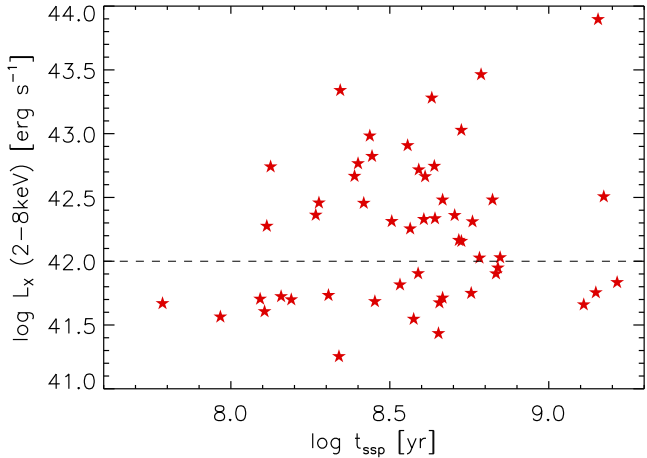}
\end{center}
\caption[]{rest-frame hard X-ray (2--8 keV) luminosity (not corrected for absorption) versus light weighted average stellar age for the XAGN sample. The dashed line shows the luminosity threshold used to compare the stellar ages of moderate luminosity and low luminosity AGN.\label{tssp-LX}}
\end{figure}

The luminosity and accretion rate of powerful AGN (L$_X$$>$10$^{44}$ erg s$^{-1}$) is found to correlate with the SFR in the host galaxy \citep[e.g.][]{Shi07,Chen13}. However, there is no clear correlation between global star formation and nuclear activity in low and moderate luminosity AGN \citep{Shao10,Mullaney12,Santini12,Rosario12,Rosario13}.
In star-forming galaxies, a recent burst of star formation can easily dominate the UV output of the galaxy, making stellar ages based on the U-V colour or \Dn appear younger even if young stars only represent a small fraction of the total stellar mass in the galaxy.
As a consequence, there is a correlation between the specific SFR and $t_{ssp}$ in star-forming galaxies (Hern\'an-Caballero et al., in prep).

Figure \ref{tssp-LX} shows the X-ray luminosity versus $t_{ssp}$ for the sources in the XAGN sample. We find no correlation between the two magnitudes. However, the concentration of sources at $\log$ $t_{ssp}$$\sim$8.6 (400 Myr) found in \S\ref{stellar-ages-subsec} seems to be stronger for AGN with L$_X$$>$10$^{42}$ erg s$^{-1}$ compared to less luminous ones. 20 out of 32 XAGN sources with L$_X$$>$10$^{42}$ erg s$^{-1}$ have 8.4$<$$\log$ $t_{ssp}$$<$8.8 compared to only 8 out of 21 with L$_X$$<$10$^{42}$ erg s$^{-1}$. The Fisher exact test gives a p-value of 0.099, which implies the difference is significant only at the $\sim$90\% confidence level. If confirmed in larger AGN samples, this would be consistent with the picture of AGN reaching peak luminosity a few hundred Myr after the onset of the starburst \citep{Davies07,Wild10}.
 
\section{Discussion}\label{sect-discussion}

The main result of this work is the existence of a significant excess of X-ray selected AGN in galaxies with intermediate stellar ages (0.3$<$$t_{ssp}$$<$0.5 Gyr), which correspond to \Dnu$\sim$1.4, after the stellar mass and redshift dependences of the AGN fraction have been accounted for.
This excess at intermediate ages is in contrast to previous results that suggested either an increased AGN fraction among the galaxies with younger populations \citep{Silverman09}, in red ones \citep{Bongiorno12} or no dependence at all \citep{Rosario13}.
While our AGN sample is small, the excess is statistically significant mostly because the procedure used to estimate stellar ages allows us to break the age-extinction degeneracy, therefore eliminating the dispersion that extinction introduces in rest-frame colours such as (U-V)$_r$.
 
A source of uncertainty for the analysis presented here is conceivably the impact of AGN emission in the measurement of stellar ages and stellar masses. While in \S\ref{sample-selection-sec} we argued that the impact of AGN emission in optical colours and derived stellar masses is negligible for galaxies with a clear 1.6\um stellar bump, it is reassuring to test whether our results can be interpreted as a consequence of contamination from the AGN emission. 
The observed distribution of \Dn for the XAGN sources can be accurately reproduced in the comparison sample of inactive galaxies by decreasing by 10\% the \Dn value of all galaxies with \Dnu$>$1.45. 
To check whether emission from a type 1 AGN in a quiescent galaxy could decrease the \Dn index by the required amount while avoiding being identified in the ACS images, we calculate the fractional AGN contribution to the rest-frame $U$ band luminosity required to reduce \Dn by 10\%.
We use the SDSS quasar composite template from \citet{VandenBerk01} as the type 1 AGN template, and the elliptical template from \citet{Coleman80} as the quiescent galaxy template. We obtain that a 10\% reduction in \Dn from AGN contamination alone requires the AGN to contribute $\sim$50\% of the integrated luminosity in the rest-frame $U$ band. 
If the AGN SED is redder or the galaxy SED is bluer than our assumption, then the required AGN contribution to the rest-frame $U$ band emission is even higher. 
We obtain conservative upper limits on the AGN contribution to the restframe $U$ band emission using ACS photometry in the $V_{606}$ band from the version r2.0z of the GOODS ACS multi-band source catalogs \citep{Giavalisco04}. Following \citep{Silverman08}, we calculate the ratio between the flux contained in circular apertures with radii 0.088" and 1". The former contains 50\% of the flux for an unresolved point source, while the later matches the aperture used for our SHARDS photometry.
We find that the median flux ratio is 4\%, which implies the AGN contributes less than 8\% to the restframe $U$ band emission in the SHARDS photometry for 50\% of the XAGN sources. The flux ratio is below 10\% for 90\% of the sample, and the highest value is 14.5\%, which represents a maximum AGN contribution $<$30\%.

AGN emission can also bias the distribution of (U-V)$_r$ and \Dn via overestimation of stellar masses. The additional emission in the NIR arising from the AGN boosts stellar mass estimates, and as a consequence AGN hosts are compared to inactive galaxies that are actually slightly more massive. The stellar mass-age correlation then makes AGN hosts seem younger. 
We use observed IRAC colours to estimate the fraction of NIR emission that arises from the AGN in the XAGN sample. We define the colour excess $\Delta$([3.6]-[5.8]) of a source as the difference between its observed [3.6]-[5.8] colour index and the average value for inactive galaxies at the same redshift. The mean $\Delta$([3.6]-[5.8]) for XAGN sources is 0.1 magnitudes. This implies the AGN contributes between $\sim$15\% and $\sim$25\% of the observed 5.8\um flux density, depending on the redshift, and less than 10\% at the peak of the stellar emission. Such a small contribution is unlikely to have any noticeable impact in stellar mass estimates.

Finally, we note that while the strongest discrepancy between distributions for the XAGN and comparison samples is found in the extinction corrected parameters ((U-V)$_0$, $t_{ssp}$) the signal is sufficiently strong in the \Dn index to rule out the extinction correction procedure as a probable cause for the observed trends. Furthermore, the lower fractions of XAGN in the quiescent locus of Figure \ref{UVJ-plot} compared to inactive galaxies confirms that AGN are more likely to be hosted in star-forming galaxies.

\section{Summary and Conclusions}\label{sect-conclusions}

We analyse the stellar populations in the host galaxies of 53 X-ray selected moderate luminosity (L$_X$$<$10$^{44}$ erg s$^{-1}$) optically faint AGN at 0.34$<$$z$$<$1.07 in the area of the GOODS-N field covered by the SHARDS survey. The ultra-deep ($m_{AB}$$<$26.5) optical medium-band (R$\sim$50) photometry from SHARDS allows us to consistently measure the strength of the 4000 \AA{} break. This, in conjunction with the rest-frame (U-V) colour, provides a 
robust measurement of the extinction that is independent of assumptions on the metallicity and SFH of the galaxies. This allows us to obtain extinction-corrected (U-V) colours and
light-weighted average stellar ages ($t_{ssp}$). 

We confirm a steep increase in the frequency of AGN with the stellar mass of an order of magnitude between 10$^{10}$ and 10$^{11}$ M$_\odot$. 50\% of our X-ray selected AGN are in hosts more massive than 10$^{11}$ M$_\odot$ and $\sim$95\% have M$_*$$>$10$^{10}$ M$_\odot$. 

A careful selection of random control samples of inactive galaxies allows us to remove the stellar mass and redshift dependences of the AGN fraction to explore trends with stellar age. We confirm that X-ray selected AGN hosts have rest-frame U-V colours comparable to those of inactive galaxies at the same mass and redshift. In particular, 2/3 of the AGN hosts in our sample and a comparable fraction of inactive galaxies are in the red sequence. However, we find that the fraction of AGN hosts with UVJ colours in the quiescent locus is only half the fraction found in inactive galaxies. The other half are instead dusty star-forming galaxies with bluer extinction-corrected colours.

\Dn measurements and extinction-corrected U-V colours both support significantly younger stellar populations in the AGN hosts, with a strong deficit of AGN among galaxies with older ($t_{ssp}$$>$1 Gyr) stellar populations. We find that X-ray detected moderate luminosity AGN ($\log$(L$_X$/erg s$^{-1}$)$\sim$41.5--44.0) are more prevalent in galaxies with intermediate stellar ages (0.3$<$$t_{ssp}$$<$0.5 Gyr) compared to younger or older galaxies, consistent with a delayed onset of AGN activity after a star formation episode. 
 
\section*{Acknowledgements}

We thank the anonymous referee for their useful comments that helped to improve this paper. A.H.-C. and A.A.-H. acknowledge funding by the Universidad de Cantabria Augusto Gonz\'alez Linares program and the Spanish Plan Nacional de Astronom\'ia y Astrof\'isica under grant AYA2012-31447. P.E. and P.G.P.-G. acknowledge support from the Spanish Plan Nacional grant AYA2012-31277. This work has made use of the Rainbow Cosmological Surveys Database, which is operated by the Universidad Complutense de Madrid (UCM), partnered with the University of California Observatories at Santa Cruz (UCO/Lick, UCSC). Based on observations made with the GTC, installed at the Spanish Observatorio del Roque de los Muchachos of the Instituto de Astrof\'isica de Canarias, in the island of La Palma.

\onecolumn

\begin{deluxetable}{lccc} % Frequency of AGN as a function of stellar age, Mass
\tabletypesize{\small}
\tablecaption{Frequency of AGN (L$_X$ [2--8keV] $>$ 10$^{41.0}$ erg s$^{-1}$)\label{XAGN-freq-table}}
\tablewidth{0pt}
\tablehead{\colhead{} & \colhead{10.0$<$$\log$(M$_*$/M$_\odot$)$<$10.5} & \colhead{10.5$<$$\log$(M$_*$/M$_\odot$)$<$11.0} & \colhead{11.0$<$$\log$(M$_*$/M$_\odot$)$<$11.5}}
\startdata
(U-V)$_r<$1.5 &  2\% (5/213) &  5\% (2/40) &  0\% (0/2)\\
1.5$<$(U-V)$_r<$2.0 &  2\% (3/161) & 10\% (17/165) & 16\% (8/50)\\
(U-V)$_r>$2.0 &  3\% (2/66) &  4\% (5/132) & 10\% (8/84)\\
\\
\hline
\\
(U-V)$_0<$0.8 &  3\% (5/185) & 13\% (7/56) & 25\% (1/4)\\
0.8$<$(U-V)$_0<$1.3 &  3\% (3/105) & 12\% (13/105) & 30\% (10/33)\\
(U-V)$_0>$1.3 &  3\% (2/60) &  3\% (4/119) &  7\% (5/69)\\
\\
\hline
\\
D$_n$(4000)$<$1.3 &  3\% (7/235) &  9\% (8/91) & 30\% (3/10)\\
1.3$<$D$_n$(4000)$<$1.5 &  3\% (2/75) & 14\% (14/103) & 25\% (11/44)\\
D$_n$(4000)$>$1.5 &  2\% (1/41) &  2\% (2/86) &  4\% (2/52)\\
\\
\hline
\\
$t_{ssp}$ $<$ 0.3 Gyr &  3\% (7/244) &  9\% (9/98) & 23\% (3/13)\\
0.3 $<$ $t_{ssp}$ $<$ 1 Gyr &  2\% (2/85) & 10\% (13/129) & 17\% (11/64)\\
$t_{ssp}$ $>$ 1 Gyr &  5\% (1/21) &  4\% (2/53) &  7\% (2/29)\\
\enddata
\end{deluxetable}
\end{document}